\def\year{2021}\relax
\documentclass[letterpaper]{article} 
\usepackage{aaai21}  
\usepackage{times}  
\usepackage{helvet} 
\usepackage{courier}  
\usepackage[hyphens]{url}  
\usepackage{graphicx} 
\usepackage{natbib}  
\usepackage{caption} 
\usepackage{booktabs} 
\usepackage{enumitem}
\usepackage{multirow}
\usepackage{diagbox}
\usepackage{cleveref}

\newcommand{\header}[1]{{\noindent{\textbf{#1}}}}

\usepackage{array} 
\usepackage[]{algorithm2e}

\title{Cross-Partisan Discussions on YouTube:\\Conservatives Talk to Liberals but Liberals Don't Talk to Conservatives}
\author{
Siqi Wu\textsuperscript{\rm 1,2}, 
Paul Resnick\textsuperscript{\rm 1} \\
}
\affiliations {
\textsuperscript{\rm 1} University of Michigan, 
\textsuperscript{\rm 2} Australian National University \\
\{siqiwu, presnick\}@umich.edu
}

\begin{document}
\maketitle

\begin{abstract}
    We present the first large-scale measurement study of cross-partisan discussions between liberals and conservatives on YouTube, based on a dataset of 274,241 political videos from 973 channels of US partisan media and 134M comments from 9.3M users over eight months in 2020.
    Contrary to a simple narrative of echo chambers, we find a surprising amount of cross-talk: most users with at least 10 comments posted at least once on both left-leaning and right-leaning YouTube channels.
    Cross-talk, however, was not symmetric.
    Based on the user leaning predicted by a hierarchical attention model, we find that conservatives were much more likely to comment on left-leaning videos than liberals on right-leaning videos.
    Secondly, YouTube's comment sorting algorithm made cross-partisan comments modestly less visible; for example, comments from conservatives made up 26.3\% of all comments on left-leaning videos but just over 20\% of the comments were in the top 20 positions.
    Lastly, using Perspective API's toxicity score as a measure of quality, we find that conservatives were not significantly more toxic than liberals when users directly commented on the content of videos.
    However, when users replied to comments from other users, we find that cross-partisan replies were more toxic than co-partisan replies on both left-leaning and right-leaning videos, with cross-partisan replies being especially toxic on the replier's home turf.
\end{abstract}

\section{Introduction}
\label{sec:intro}

Political scientists and communication scholars have raised concerns that online platforms can act as echo chambers, reinforcing people's pre-existing beliefs by exposing them only to information and commentary from other people with similar viewpoints~\cite{iyengar2009red,flaxman2016filter}.
Echo chambers are concerning because they can drive users to adopt more extreme positions and reduce the chance for building the common ground and legitimacy of political compromises~\cite{pariser2011filter,sunstein2018republic}.
While prior research has investigated the (negative) effects of echo chambers based on small-scale user surveys and controlled experiments~\cite{an2014partisan,dubois2018echo}, one fundamental question still remains unclear: to what degree do users experience echo chambers in real online environments?

One countermeasure for mitigating echo chambers is to promote information exchange among individuals with different political ideologies.
Cross-partisan communication, or cross-talk, has thus become an important research subject.
Researchers have studied cross-partisan communication on platforms such as Twitter~\cite{lietz2014politicians,garimella2018quantifying,eady2019many}, Facebook~\cite{bakshy2015exposure}, and Reddit~\cite{an2019political}.
However, little is known about YouTube, which has become an emerging source for disseminating news and an active forum for discussing political affairs.
According to a recent survey from Pew Research Center~\cite{stocking2020many}, 26\% Americans got news on YouTube, and 51\% of them primarily looked for opinions and commentary on the videos.

\begin{figure}[tbp]
    \centering
	\includegraphics[width=0.99\columnwidth]{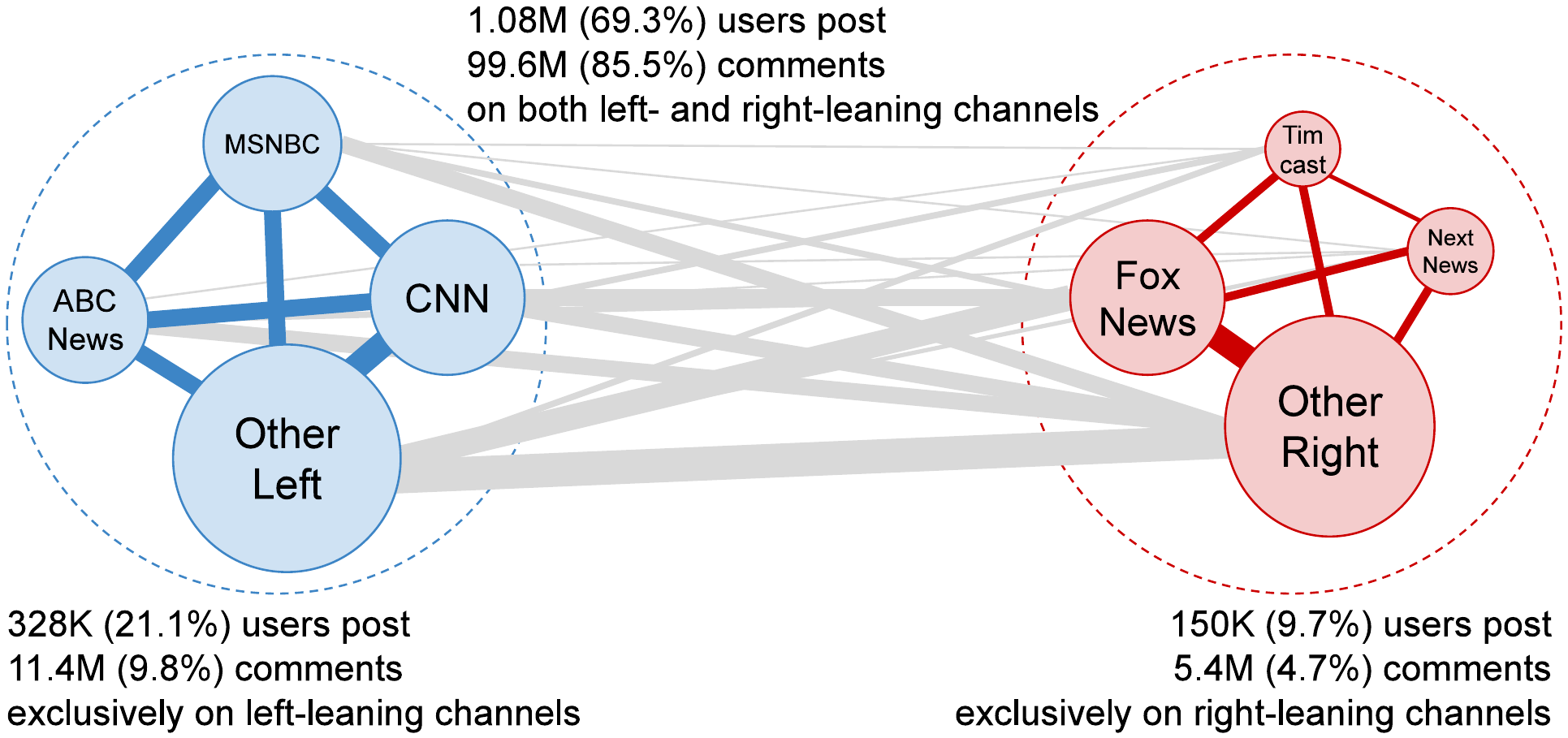}
	\caption{There is a surprising amount of cross-talk between left-leaning and right-leaning YouTube channels.
	Each node is a channel.
	Node size is proportional to the number of users who comment.
	Edge width is scaled by the number of users commenting on channels at both ends.
    This plot reflects only active users, those who posted at least 10 comments.
    }
	\label{fig:intro-teaser}
\end{figure}

To fill this gap, we curated a new dataset that tracked user comments on YouTube videos from US political channels.
Focusing on active users who posted at least 10 comments each, the top-line surprising result is that most commenters did not confine themselves to just left-leaning or just right-leaning channels\footnote{We replicated the analyses based on users with at least 2 comments to users with at least 30 comments. The results are qualitatively similar and supplemented in~\Cref{sec:si_cp_user}.}.
\Cref{fig:intro-teaser} visualizes the shared audience network between channels.
The blue and red dashed circles respectively enclose the left-leaning and right-leaning channels. 
In each circle, we show the top 3 most commented channels.
Edges represent audience overlap, i.e., the number of people who commented on both channels.
We find that 69.3\% of commenters posted at least once on both left-leaning and right-leaning channels.
Together, those users' comments made up 85.5\% of all comments.
This example illustrates that a significant amount of YouTube users have participated in cross-partisan political discussions.

In this work, we tackled three open questions related to cross-partisan discussions on YouTube.
First, \textbf{what is the prevalence of cross-talk?}
The theory of selective exposure suggests that people deliberately avoid information challenging their viewpoints~\cite{hart2009feeling}.
Thus, one might expect few interactions across political lines.
However, survey studies have found conflicting evidence that people are aware of and sometimes even seek out counter-arguments~\cite{horrigan2004internet,an2014partisan}, and that
political filter bubbles are actually occurring to a limited extent~\cite{an2011media,dubois2018echo}.
To address this question, we trained a hierarchical attention model~\cite{yang2016hierarchical} to predict a user's political leaning based on the comment collection s/he posted.
We find that cross-partisan communication was not symmetric: on the user level, conservatives were more likely to post on left-leaning videos than liberals on right-leaning videos;
on the video level, especially on videos published by right-leaning independent media, there were relatively few comments from liberals.

Second, \textbf{does YouTube promote cross-talk?}
YouTube's recommender system has been criticized for creating echo chambers in terms of which videos users would be recommended~\cite{lewis2018alternative}.
Here, we examined whether the sorting of comments on the video pages tended to suppress cross-talk.
We examined the fraction of cross-partisan comments shown in the top 20 positions.
While the position bias for each successive position was small, overall comments from the opposite ideologies appeared less frequently among the top positions than their total numbers would predict.

Third, \textbf{what is the quality of cross-talk?}
Discussions on YouTube are multifaceted -- a user can reply to the content of videos, or reply to another user who commented before.
The conversation quality may differ depending on whom the comments reply to and where the discussions take place.
We used toxicity as a measure of quality and used an open tool -- Perspective API~\cite{jigsaw2020} -- to classify whether a comment is toxic or not.
We find only small differences in toxicity of comments targeting the video content.
We find comments that reply to people with opposing views were much more frequently toxic than co-partisan replies.
Furthermore, cross-partisan replies were especially more toxic on a replier's home territory: i.e., when liberals replying to conservatives on left-leaning videos and when conservatives replying to liberals on right-leaning videos.

The main contributions of this work include:
\begin{itemize}[leftmargin=*]
    \item a procedure for estimating the prevalence of cross-partisan discussions on YouTube;
    \item a finding that cross-partisan comments are common, but much more frequently by conservatives on left-leaning videos than by liberals on right-leaning videos;
    \item a finding that cross-partisan comments are less likely to appear in the top positions on the video pages;
    \item a finding that cross-partisan comments are more toxic, and further deteriorate when replying to users of the opposite ideology on one's home turf;
    \item a new political discussion dataset that contains 274,241 political videos published by 973 channels of US partisan media and 134M comments posted by 9.3M users.\footnote{The code and anonymized data are publicly available at \url{https://github.com/avalanchesiqi/youtube-crosstalk}.
    }
\end{itemize}

\section{Related work}
\label{sec:related}

\subsection{Echo chambers and affective polarization}

Research on echo chamber dates back to cognitive dissonance theory in the 1950s~\cite{festinger1957theory}, which argued that people tended to seek out information that they agreed on in order to minimize dissonance.
Since then, there have been persistent concerns that the advances of modern techniques may exacerbate the extent of ideological echo chambers by making it easier for users to receive a personalized information feed that \citet{negroponte1996being} referred to as the ``Daily Me'' or even be automatically insulated inside the ``filter bubbles'' by algorithms that feed users more things like what they have seen before~\cite{pariser2011filter}. Concerns have been raised that isolation in echo chambers will drive people towards more extreme opinions over time~\cite{sunstein2018republic} and to heightened affective polarization -- animosity towards people with different political views.

However, evidence is mixed on the extent to which people really prefer to be in the echo chambers. \citet{iyengar2009red} found partisan divisions in attention that people gave to articles based on the ideological match with the mainstream news source (Fox News, NPR, CNN, BBC) to which articles were artificially attributed to. On the other hand, \citet{garrett2009echo} argued that people have a preference for encountering some reinforcing information but not necessarily an aversion to challenging information, and \citet{munson2010presenting} found that some people showed a preference for collections of news items that had more ideological diversity. \citet{gentzkow2011ideological} found that ideological segregation of online news consumption was relatively low and less than ideological segregation of face-to-face interactions, but higher than segregation of offline news consumption. \citet{dubois2018echo}, in a survey of 2,000 UK Internet users, found that people who were more politically engaged reported more exposure to opinions that they disagreed with and more exposure to things that might change their minds. In a case study, \citet{quattrociocchi2016echo} found little overlap between the users who commented on Facebook posts articulating conspiracy narratives vs. on mainstream science narratives under the same topic. In a study based on a sample of Twitter users, \citet{eady2019many} found that the most conservative fifth had more than 12\% of the media accounts that they followed at least as left as the New York Times; the liberal fifth had 4\% of the media accounts that they followed at least as right as Fox News.

Evidence is also mixed on the extent to which algorithms are promoting echo chambers, and whether they lead people to more extreme opinions. \citet{an2011media} found that many Twitter users were indirectly exposed to media sources with a different political ideology. In a large-scale browsing behavior analysis, \citet{flaxman2016filter} found that people were less segregated in the average leaning of media sites that they accessed directly than those they accessed through search and social media feeds, but there was more variety in the ideologies of individuals who were exposed through search and social media.
On Facebook, \citet{bakshy2015exposure} found substantial exposure to challenging viewpoints; more than 20\% of news articles that liberal clicked on was cross-cutting (meaning it had a primarily conservative audience) and for conservatives, nearly 30\% of news articles that they clicked on had a primarily liberal audience.
\citet{hussein2020measuring} audited five search topics on YouTube and found that watching videos containing misinformation led to more misinformation videos being recommended to watch. \citet{ribeiro2020auditing} found that users migrated over time from alt-lite channels to more extreme alt-right channels, but that the video recommender algorithm did not link to videos from those more extreme channels; \cite{ledwich2020algorithmic} also found that the YouTube recommender pushed viewers toward mainstream media rather than extreme content.

Finally, there have also been studies with conflicting results about whether exposure to diverse views increases or decreases affective polarization. \cite{garrett2014implications} hypothesized that exposure to concordant news sources would activate partisan identity and thus increase affective polarization, while exposure to out-party news sources will reduce partisan identity and thus affective polarization. They found evidence consistent with these hypotheses based on surveys of U.S. and Israeli voters. On the other hand, in a field experiment, \citet{bail2018exposure} found that people who were paid to follow a bot account that retweeted opposing politicians and commentators became more extreme in their own opinions, though it did not directly assess changes in affective polarization. 

\subsection{The nature of cross-talk}

The impact of exposure to opposing opinions may well depend on how that information is presented. This is especially true for direct conversations between people.
There is a huge difference between a respectful conversation on the "/r/changemyview" subreddit and having a troll join a conversation with the goal of trying to disrupt or elicit angry reactions~\cite{phillips2015we, flores2018mobilizing}. \citet{bail2021} (Chapter 9) described an experiment where encounters with others of opposing views were structured to hide some ideological and demographic characteristics; this led to more moderate political views and reduced affective polarization in a survey conducted a week later.

Some but not all cross-cutting political discussion online is adversarial or from trolls. \citet{hua2020towards} developed a technique for measuring the extent to which interactions with a particular political figure were indeed adversarial. \citet{an2019political} found that Reddit users used more complex language and sophisticated reasoning, and more posing of questions in two cross-cutting subreddits than those same users did in two homogenous subreddits. More broadly, \cite{rajadesingan2020quick} reported that across a large number of political subreddits, most individuals who participated in multiple subreddits adjusted their toxicity levels to more closely match that of the subreddits they participated in.

\subsection{Estimating user political ideology}

Semi-supervised label propagation is a popular method for predicting user political leaning~\cite{zhou2011classifying,cossard2020falling}.
The core assumption of label propagation is homophily -- actions that form the base network must be either endorsement or refutation, but not both.
However, commenting is ambiguous, which may express attitudes of both agreement and disagreement.
Some researchers developed Monte Carlo sampling methods for clustering users~\cite{barbera2015birds,garimella2018quantifying}, while others tried to extract predictive textual features~\cite{preoctiuc2017beyond,hemphill2020two}.
Recent advancements in language modeling allow us to infer user leaning directly from text embedding~\cite{yang2016hierarchical}.
In this work, we implemented a hierarchical attention network model for predicting user political leaning.

\section{YouTube political discussion dataset}
\label{sec:dataset}

We constructed a new dataset by collecting user comments on YouTube political videos.
Our dataset consists of 274K videos from 973 US political channels between 2020-01-01 and 2020-08-31, along with 134M comments from 9.3M users.
In this section, we first describe the collection of YouTube channels of US partisan media.
We then introduce our scheme for coding these channels by types such as national, local, or independent media.
Lastly, we describe our procedure for collecting videos and comments.

\subsection{YouTube channels of US partisan media}
\label{ssec:data-media}

\header{MBFC-matched YouTube channels.}
We scraped the bias checking website Media Bias/Fact Check\footnote{\url{https://mediabiasfactcheck.com/}} (MBFC), which is a commonly used resource in scientific studies~\cite{bovet2019influence,ribeiro2020auditing}.
This yielded 2,307 websites with reported political leanings.
MBFC classifies the leanings on a 7-point Likert scale (extreme-left, left, center-left, center, center-right, right, extreme-right).
The exact details of their procedure are not described, but the process takes into account word choices, factual sourcing, story choices, and political affiliation.
Since our focus was on the extent of cross-partisan interactions, we collapsed this 7-point classification: 
we collapsed \{extreme-left, left, center-left\} and \{extreme-right, right, center-right\} each into one left-leaning and one right-leaning group.
We discarded 186 center media.
Additionally, the first author examined the articles and ``about'' page of each site to annotate the media's country.
Because the ratings of MBFC were based on the US political scale, we only kept the 1,410 US websites.

Next, we scraped all those websites to detect YouTube links.
This yielded 450 websites with YouTube channels.
For the unresolved websites, we queried the YouTube search bar with their URLs.
We crawled the channel ``about'' pages of the top 10 search results and retained the first channel with a hyperlink redirecting to the queried website.
To improve the matching quality, we used the \texttt{SequenceMatcher} Python module to compute a similarity score between the YouTube channel title and website title.
The score ranges from 0 to 1, where 1 indicates exact match.
The first author manually checked and labeled any site with either similarity score below 0.5 or with no search result.
In total, this mapped 929 YouTube channels to the corresponding US partisan websites, 640 of which published at least one video in 2020.

\header{Addition of channels featured on those channels.}
The MBFC website has good coverage of mainstream media and local newspapers, but not of YouTube political commentators.
Studies have found that scholars, journalists, and pundits regularly use YouTube to promote political positions~\cite{lewis2018alternative}.
To expand our dataset, we crawled the ``featured channels'' listed on the 640 MBFC-matched active YouTube channels.
Channels choose which other channels they want to feature in this way.
We did not observe any channel that was featured by both a left-leaning channel and a right-leaning channel among our 640 channels.
Therefore, we added these featured channels to our dataset and labeled them with the same political ideology as the channels featuring it.
Our approach was validated on the 62 MBFC-matched channels that were also featured by some other MBFC-matched channels.
We compared their predicted leanings to the MBFC labels and found that all 62 channels are correctly predicted.
This yielded 637 additional channels.

\header{Addition of existing YouTube media bias datasets.}
We further augmented our list of channels with three existing YouTube media bias datasets~\cite{lewis2018alternative,ribeiro2020auditing,ledwich2020algorithmic}.
These datasets were annotated by the authors of corresponding papers.
After discarding the intersections and center channels, we obtained 356 additional channels with known political leanings.

\subsection{Channel coding scheme}

The first author manually coded the collected channels by assessing their descriptions on YouTube and Wikipedia.
Pew Research Center has also proposed a similar coding scheme~\cite{stocking2020many}.

\begin{itemize}[leftmargin=*]
    \item National media: channels of televisions, radios, or newspapers from major US news conglomerates (e.g., CNN, Fox News, New York Times) and high profile politicians (e.g., Donald Trump and Joe Biden). They cover stories across multiple areas and have national readership.
    \item Local media: channels of media publishing or broadcasting in small finite areas. They usually focus on local news and have local readership (e.g., Arizona Daily Sun, Texas Tribune, ABC11-WTVD, CBS Boston).
    \item Organizations: channels of governments, NGOs, advocacy groups, or research institutions (e.g., White House, GOP War Room, Hoover Institution).
    \item Independent media: channels that have no clear affiliation with any organization (e.g., The Young Turks, Breitbart) and content producers who make political commentary videos (e.g., Ben Shapiro, Paul Joseph Watson).
\end{itemize}

\subsection{Video and comment collection}

\begin{table}
    \centering
    \small
    \begin{tabular}{rrrrrr}
        \toprule
         & left-leaning & right-leaning & total \\
        \midrule
        \#channels & 462 (47.5\%) & 511 (52.5\%) & 973  \\
        \#videos & 195,482 (71.3\%) & 78,759 (28.7\%) & 274,241 \\
        \#views & 13.4B (71.7\%) & 5.3B (28.3\%) & 18.7B \\
        \#comments & 79.5M (59.4\%) & 54.3M (40.6\%) & 133.8M \\
        \bottomrule
    \end{tabular}
    \caption{Statistics of YouTube political discussion dataset.}
    \label{table:videos}
\end{table}

\header{Filtering out non-political channels.}
We removed non-political channels (e.g., music, gaming) because comments by liberals and conservatives on these channels are often not about politics and thus are not necessarily indicative of cross-partisan communication.
On the channel level, since YouTube did not provide a category label, the first author manually reviewed 10 random videos of each channel from 2020.
Channels were removed if they did not publish any video in 2020 or none of their 10 randomly selected videos was relevant to US news and politics.

\header{Filtering out non-political videos.}
Political channels sometimes published non-political videos, which we wanted to exclude from our dataset.
Channels can tag each video with one of 15 categories.
We kept videos that were tagged with the following six categories: ``News \& Politics, Nonprofits \& Activism, Entertainment, Comedy, People \& Blogs, Education''.
Categories such as Music, Sport, and Gaming were excluded.
We retained Entertainment and Comedy because we found that independent media often used these tags even when the videos were clearly related to politics.  

\header{Collecting video and comment metadata.}
Altogether, we collected 274,241 political videos where commenting was permitted, and published by the 973 channels of US partisan media between 2020-01-01 and 2020-08-31.
For each video, we collected its title, category, description, keywords, view count\footnote{Our data collection started on 2020-10-15 and took a week to finish. All collected videos had been on YouTube for at least 45 days. A previous study of 459,728 political videos from a public YouTube dataset~\cite{wu2018beyond} found that on average, 90.6\% of views were accumulated within the first 45 days. Thus, we considered our observed video view counts as the ``lifetime'' view counts.}, transcripts, and comments.

\begin{figure}[tbp]
    \centering
	\includegraphics[width=0.99\columnwidth]{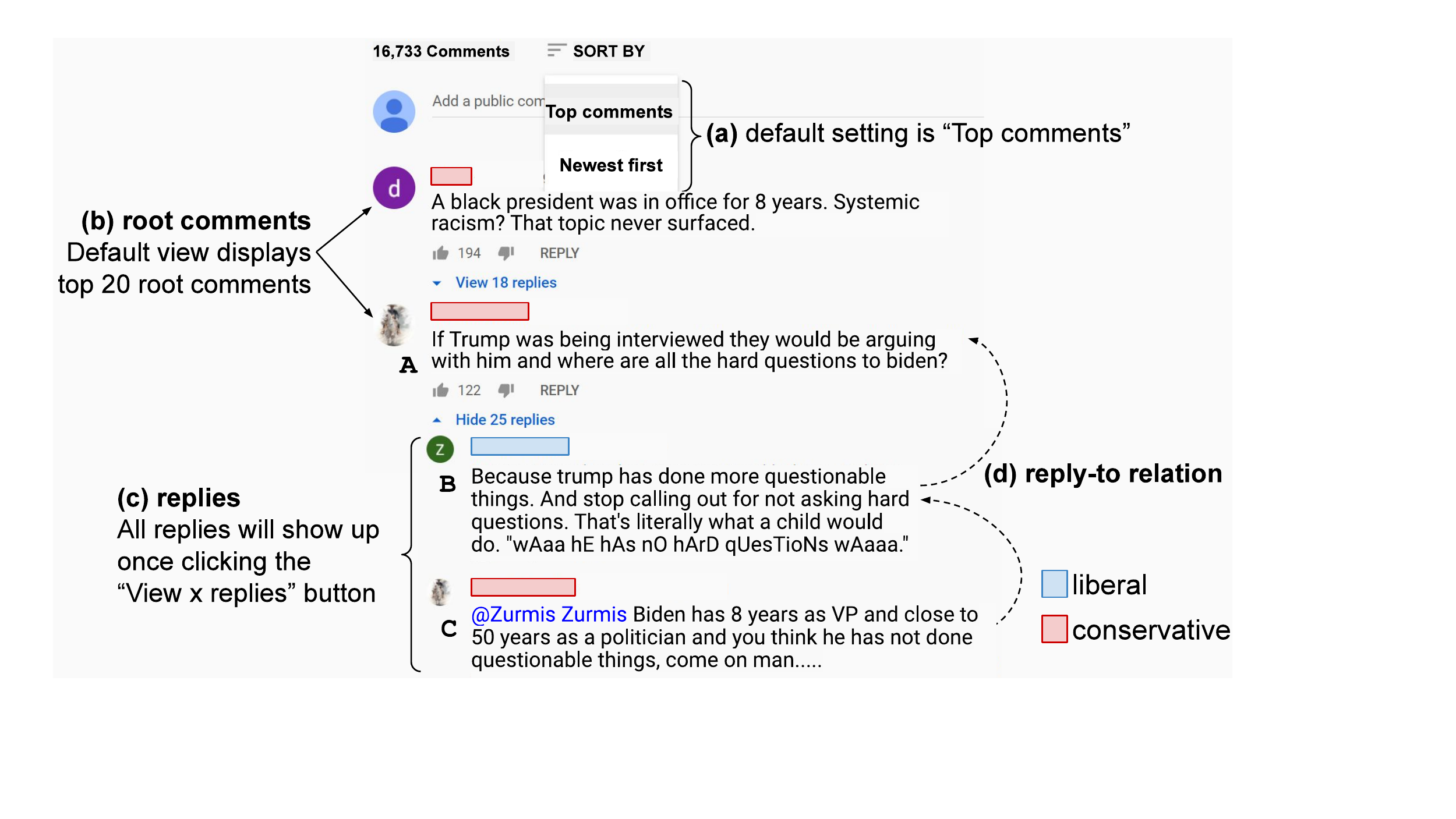}
	\caption{Snapshot of the comment section of a CNN video (id: \texttt{B2e35AbLP\_Y}).
	(a) two options to sort comments.
	(b) root comments. 
	(c) replies under a root comment.
	(d) reply-to relation between comments, e.g., B replied to A, and C replied to B.
	Usernames were shaded by their predicted political leanings.
	The two highest ranked root comments on this left-leaning CNN video were from conservatives.
	}
	\label{fig:data-webpage}
\end{figure}

YouTube provided two options for sorting video comments -- \texttt{Top comments} or \texttt{Newest first} (\Cref{fig:data-webpage}a).
Without scrolling or clicking, the YouTube web interface displayed the top 20 root comments by default (\Cref{fig:data-webpage}b), making them enjoy significantly more exposure than the collapsed replies (\Cref{fig:data-webpage}c).
We first queried the \texttt{Top comments} tab to scrape the 20 root comments for investigating the bias in YouTube's comment sorting algorithm (see~\Cref{sec:system}).
We also queried the \texttt{Newest first} tab to scrape all the root comments and replies.
For each comment, we collected its comment id, author, and text.
Note that root comments and replies had distinct formats of comment ids -- let's assume the comment id of a root comment was ``\texttt{A}'', then its replies would have comment ids of the format ``\texttt{A.B}''.
The comment ids allowed us to reconstruct the conversation threads under root comments, i.e., who replied to whom.
For example, both ``B replied to A'' and ``C replied to B'' in \Cref{fig:data-webpage}(d) were cross-partisan comments.
In total, we collected 133,810,991 comments from 9,304,653 users.

\begin{figure}[!tbp]
    \centering
	\includegraphics[width=0.99\columnwidth]{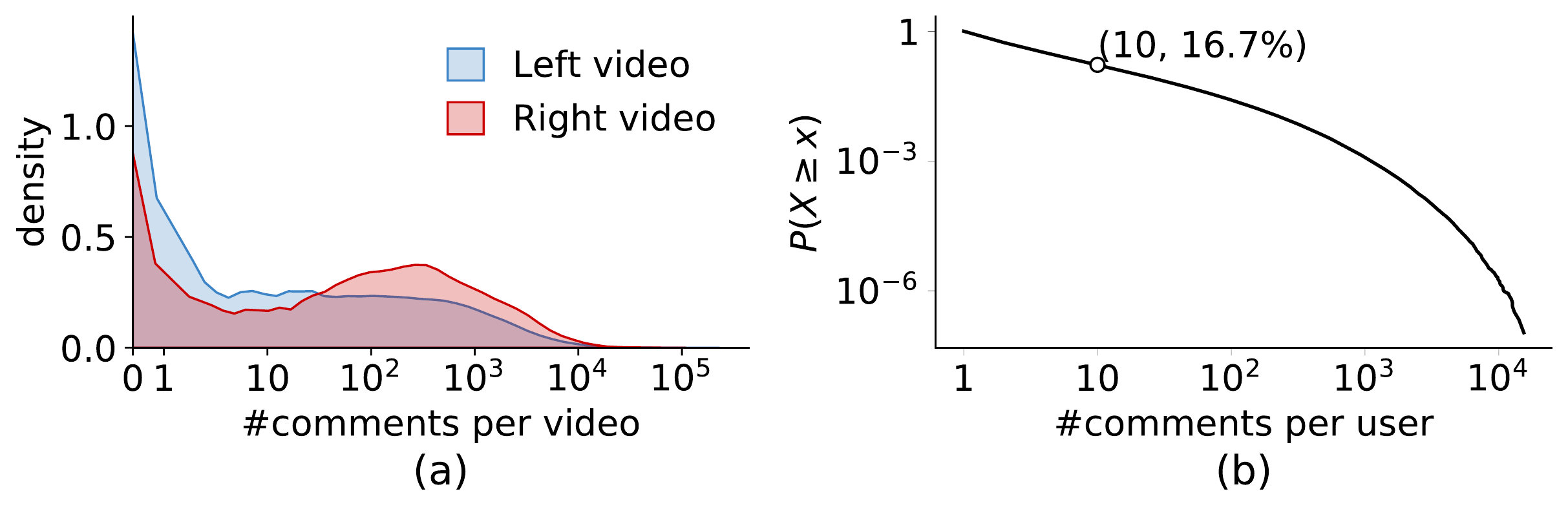}
	\caption{(a) PDF of comments per video, disaggregated by political leanings.
	(b) CCDF of comments per user.
	}
	\label{fig:data-profiling}
\end{figure}

\Cref{table:videos} summarizes the dataset statistics.
We collected more right-leaning channels but they published fewer videos on average.
Left-leaning videos had similar average views but fewer comments per video compared to right-leaning videos.
\Cref{fig:data-profiling}(a) plots the probability density function (PDF) of comments per video.
We observe a higher portion of right-leaning videos with large comment volume.
In \Cref{fig:data-profiling}(b), we plot the complementary cumulative density function (CCDF) of comments per user.
16.7\% of users posted at least 10 comments.

\section{Predicting user political leaning}
\label{sec:predict}

In this section, we first determined the ideological leaning for a set of seed users based on (a) political hashtags and URLs in the comments, and (b) left-leaning and right-leaning channels that users subscribed to.
Next, we trained a hierarchical attention model based on the comment texts from seed users, and then predicted the political leaning for the remaining users.

\subsection{Obtaining labels for seed users}
\label{ssec:predict-lp}

\header{Political hashtags and URLs.}
Hashtags are widely used to promote online activism.
Research has shown that the adoption of political hashtags is highly polarized -- users often use hashtags that support their ideologies and rarely use hashtags from the opposite side~\cite{hua2020towards}.
We extracted 231,483 hashtags from all 134M comments.
The first author manually examined the 1,239 hashtags appearing in at least 100 comments, out of which 306 left and 244 right hashtags were identified as seed hashtags\footnote{We did not include hashtags that describe a person (e.g., \#donaldtrump) or a social movement (e.g., \#metoo) as they may be used by both liberals and conservatives. Instead, we considered hashtags that posit clear stance towards political parties (e.g., \#voteblue, \#nevertrump), or hashtags that promote agendas mostly supported by one party (e.g., \#medicareforall, \#prolife).}.

Users sometimes shared URLs in the comments.
If a URL redirected to a left-leaning MBFC website or a left-leaning YouTube video, we replaced the URL text in the comment with a special token \texttt{LEFT\_URL}.
We resolved to \texttt{RIGHT\_URL} in the same way.

For simplicity, we used the term ``entity'' to refer to both hashtag and URL. 
We constructed a user-entity bipartitie graph, in which each node was either a user or an entity.
Undirected edges were formed only between users and entities.
Next, we applied a label propagation algorithm on the bipartite graph~\cite{liu2016pay}.
The algorithm was built upon two assumptions: (a) liberals (conservatives) mostly use left (right) entities; (b) left (right) entities are mostly used by liberals (conservatives).

$\triangleright$ {\em Step 1: from entities to users.}
For each user, we counted the number of used left and right entities, denoted as $E_l$ and $E_r$, respectively.
Users who shared fewer than 5 entities, i.e., $E_l + E_r < 5$, were excluded.
We used the following equation to compute a homogeneity score:

\begin{equation}
    \label{eq:purity}
    \delta(l, r) = \frac{r - l}{r + l}
  \end{equation}

$\delta$ ranges from -1 to 1, where -1 (or 1) indicates that this user exclusively used left (or right) entities.
The same scoring function has also been used by~\citet{robertson2018auditing}.
Since we focused on the precision rather than recall in identifying seed users, we considered a user to be liberal if $\delta (E_l, E_r) \leq -0.9$ and conservative if $\delta (E_l, E_r) \geq 0.9$.

$\triangleright$ {\em Step 2: from users to entities.}
We only propagated user labels to hashtags, but not URLs because most URLs apart from the MBFC websites were non-political.
For each hashtag, we counted the number of liberals and conservatives who used it, denoted as $H_l$ and $H_r$, respectively. 
We excluded hashtags shared by fewer than 5 users, i.e., $H_l + H_r < 5$.
Using \Cref{eq:purity}, a hashtag was considered left if $\delta (H_l, H_r) \leq -0.9$ and considered right if $\delta (H_l, H_r) \geq 0.9$.

Beginning with the manually identified 306 left and 244 right hashtags, we repeatedly predicted user leanings in {\em Step 1}, and then updated the set of political hashtags in {\em Step 2}.
We repeated this process until no new user nor new hashtag was obtained.
This yielded 8,616 liberals and 8,144 conservatives, denoted as $Seed_{ent}$, based on an expanded set of 834 left and 717 right hashtags.

\header{User subscriptions.}
For the 1,555,428 (16.7\%) users who posted at least 10 comments, we scraped their subscription lists on YouTube, i.e., channels that a user subscribed to.
A total of 476,701 (30.6\%) users subscribed to at least one channel.
For each user, we counted the number of left-leaning and right-leaning channels in their subscription list, denoted as $S_l$ and $S_r$, respectively.
We excluded users who subscribed to fewer than 5 partisan channels, i.e., $S_l + S_r < 5$.
Using \Cref{eq:purity}, a user was considered liberal if $\delta (S_l, S_r) \leq -0.9$ and considered conservative if $\delta (S_l, S_r) \geq 0.9$.
This yielded 61,320 liberals and 86,134 conservatives from subscriptions, denoted as $Seed_{sub}$.

\header{Validation.}
We compared $Seed_{ent}$ to $Seed_{sub}$.
They had an intersection of 2,035 users, among which 1,958 (96.2\%) users had the same predicted labels, suggesting a very high agreement.
Given this validation, we merged $Seed_{ent}$ with $Seed_{sub}$, while removing the remaining 77 users with conflicted labels.
This leaves a total of 162,102 seed users (68,883 liberals and 93,219 conservatives).

Mentioning a hashtag, sharing an URL, and subscribing to a channel are all endorsement actions.
They all satisfy the homophily assumption of the label propagation algorithms.
However, most users in our dataset did not have such endorsement behaviors.
The coverage of identified seed users was low (162K out of 9.4M, 1.7\%).
For the remaining users, we used methods that could infer user political leanings solely based on the collection of comment texts.

\subsection{Classification by Hierarchical Attention Network}
\label{ssec:predict-han}

Hierarchical Attention Network (HAN) is a deep learning model for document classification~\cite{yang2016hierarchical}.
HAN differentiates itself from other text classification methods in two ways: (a) it captures the hierarchical structure of text data, i.e., a document consists of many sentences and a sentence consists of many words; (b) it implements an attention mechanism to select important sentences in the document and important words in the sentences.
Researchers have applied HAN in many classification tasks on social media and shown that HAN is one of the top performers~\cite{cheng2019hierarchical}.
HAN is a suitable choice for our task of predicting user political leaning, because a user posts many comments and each comment consists of many words.

\header{Experiment setup.}
We used the 162K seed users as training set.
For each user, we treated their comments as independent sentences, and we concatenated all comments into one document.
We replaced URLs by their primary domains.
Punctuation and English stopwords were also removed.
We employed 5-fold cross validation, which ensured that all seed users will be estimated and evaluated once.

\begin{figure}[tbp]
    \centering
	\includegraphics[width=0.99\columnwidth]{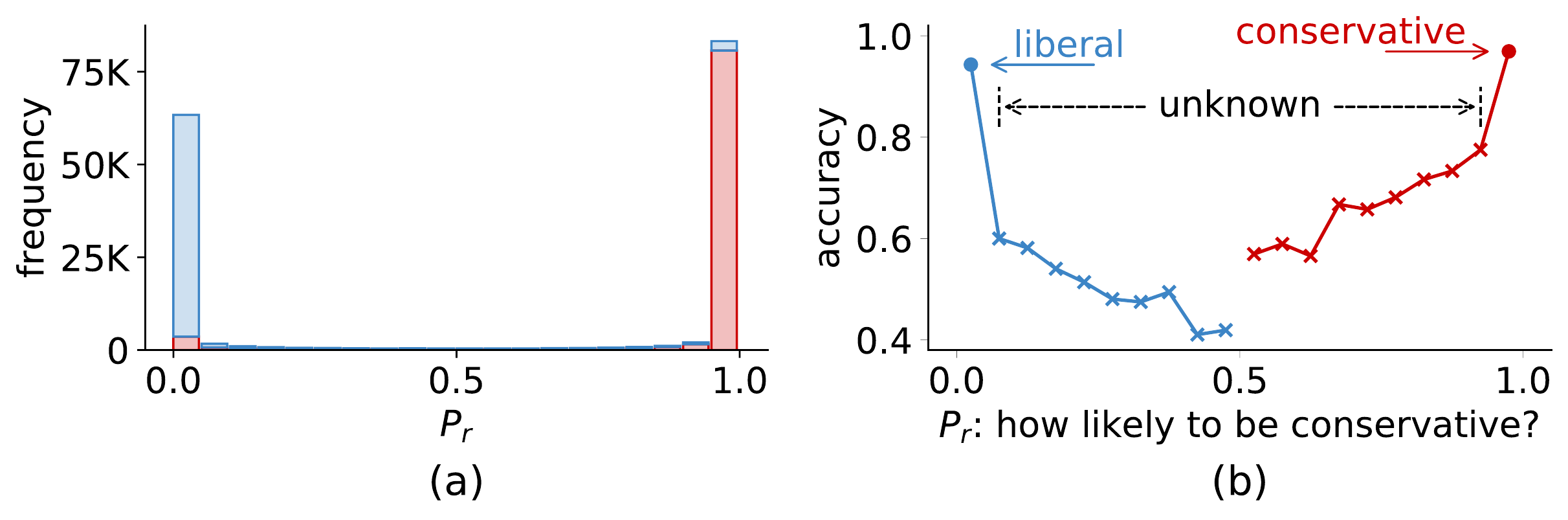}
	\caption{
	HAN classification results of 162K seed users.
	The x-axis is $P_r$ from the HAN model, which indicates how likely a user to be conservative.
	(a) The HAN model is confident (i.e., $P_r \leq 0.05$ or $P_r \geq 0.95$) in predicting the majority of seed users.
	(b) When the HAN model is confident, it achieves very high accuracy.
	We classified user political leaning as unknown if $0.05 < P_r < 0.95$.}
	\label{fig:prediction-han}
\end{figure}

\header{Prediction results.}
The outputs of HAN are two normalized values $P_l$ and $P_r$ via the softmax activation ($P_l + P_r = 1$).
The model classifies a user as conservative if $P_r > 0.5$, otherwise as liberal.
\Cref{fig:prediction-han}(a) shows the frequency distribution of $P_r$ for the 162K seed users.
We binarized $P_r$ into 20 buckets.
91.6\% of seed users fell into the leftmost ($P_r \leq 0.05$) or the rightmost ($P_r \geq 0.95$) buckets, suggesting that our trained HAN was very confident about the prediction results.
We show the number of true liberals and true conservatives in each bucket as a stacked bar.
In~\Cref{fig:prediction-han}(b), we plot the accuracy of each bucket.
The accuracy was 0.943 when predicting liberal with $P_r \leq 0.05$, and 0.969 when predicting conservative with $P_r \geq 0.95$.
The prediction power dropped significantly when $0.05 < P_r < 0.95$, but there were relatively few users in that range. 
The overall accuracy was 0.929.

In practice, some users may have no comment that can reveal their political leanings, e.g., comment spammers. 
Other users may have a mixed ideology. 
We thus divided users into three bins based on thresholds of our HAN model output: a user is classified as liberal if $P_r \leq 0.05$, as conservative if $P_r \geq 0.95$, otherwise as unknown (see~\Cref{fig:prediction-han}b).

After training, we had five HAN models, each trained on one fold of 80\% data.
In the inference phase, if any two models gave conflicting predictions for the same user, we set that user to the unknown class.
This yielded known classifications for 6,370,150 users (3,738,714 liberals and 2,631,436 conservatives).
Combining with the 162K seed users, we assigned political leanings for 6.53M (69.7\%) users, who posted 123M (91.9\%) comments.
For comparison, we also implemented a Vanilla LSTM model, which achieved similar overall accuracy of 0.922.
However, after setting users with $0.05 < P_r < 0.95$ to unknown, LSTM yielded a lower coverage of classified users.
We thus used the classification results from HAN in the following analysis.

\begin{figure}[tbp]
    \centering
	\includegraphics[width=0.99\columnwidth]{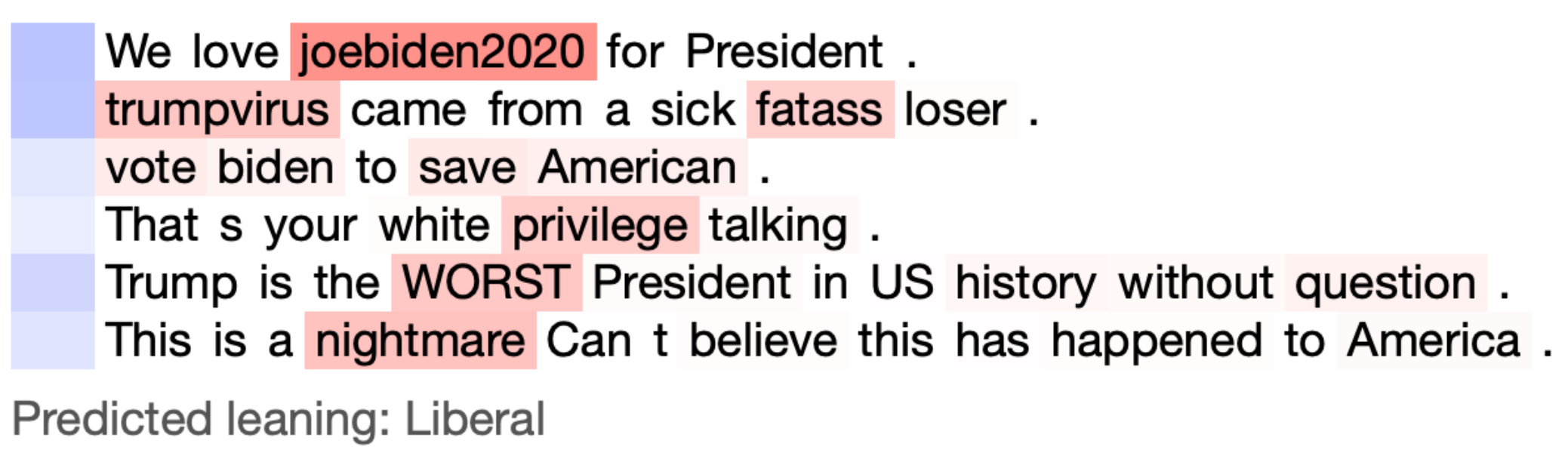}
	\caption{Visualizing HAN.
	Darker blue indicates important comments, while darker red indicates important words.}
	\label{fig:prediction-vis}
\end{figure}

\header{Visualizing HAN.}
One advantage of HAN is that it can differentiate important sentences and words.
\Cref{fig:prediction-vis} visualizes the attention masks for comments from a liberal user.
Each line is a separate comment.
The top two comments contribute the most to prediction result.
Our model can select words with correct political leaning.
Those include existing words in our expanded hashtag set such as {\em joebiden2020} and {\em trumpvirus}, as well as new words such as {\em privilege}, {\em worst}, and {\em nightmare}.

\section{Prevalence analysis}
\label{sec:prevalence}

We studied three questions related to the prevalence of cross-partisan discussions between liberals and conservatives.
We first quantified the portions of cross-partisan comments from a user-centric view, then and from a video-centric view.
Finally, we investigated whether the extent of cross-talk varied on different media types.

\subsection{How often do users post on opposing channels?}

We empirically measured how often a YouTube user commented on videos with opposite ideologies.
For active users with at least 10 comments, we counted the frequencies that they posted on left-leaning and right-leaning videos.
The HAN model classified 90.1\% of the active users as either liberal or conservative.
Among them, 62.2\% of liberals posted at least once on right-leaning videos, while 82.3\% conservatives posted at least once on left-leaning videos.
    
\begin{figure}[ht]
    \centering
	\includegraphics[width=0.99\columnwidth]{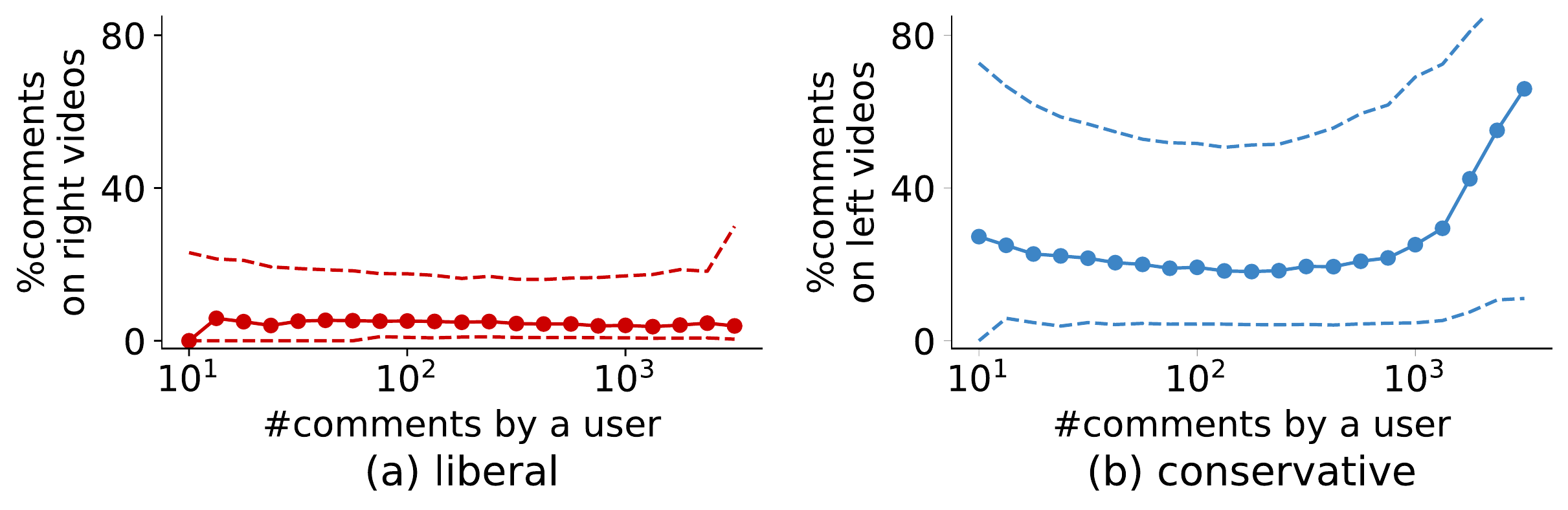}
	\caption{Analysis of users' cross-partisan interaction: conservatives were more likely to comment on left-leaning videos than liberals on right-leaning videos.
	The x-axis shows the total number of comments from the user, split into 21 equally wide bins in log scale.
	The y-axis shows the percentage of comments on videos with opposing ideologies.
	Within each bin, we computed the 1st quartile, median, and 3rd quartile.
	The line with circles connects the medians, while the two dashed lines indicate the inter-quartile range.
	}
	\label{fig:measure-user}
\end{figure}

\Cref{fig:measure-user} plots the fraction of users' comments that were cross-partisan, as a function of how prolific the users were at commenting. Overall, conservatives posted much more frequently on left-leaning videos (median: 22.2\%, mean: 33.9\%) than liberals on right-leaning videos (median: 4.8\%, mean: 15.6\%). 
The fractions of cross-partisan comments from liberals were largely invariant to user activity level (\Cref{fig:measure-user}a). 
By contrast, prolific conservatives disproportionately commented on left-leaning videos (\Cref{fig:measure-user}b). 
The few most prolific conservative commenters with more than 10,000 comments made more than half their comments on left-leaning videos, suggesting a potential trolling behavior.
Nevertheless, even for less prolific conservatives, they still commented on left-leaning videos far more frequently than liberals did on right-leaning videos.

\begin{figure}[ht]
    \centering
	\includegraphics[width=0.99\columnwidth]{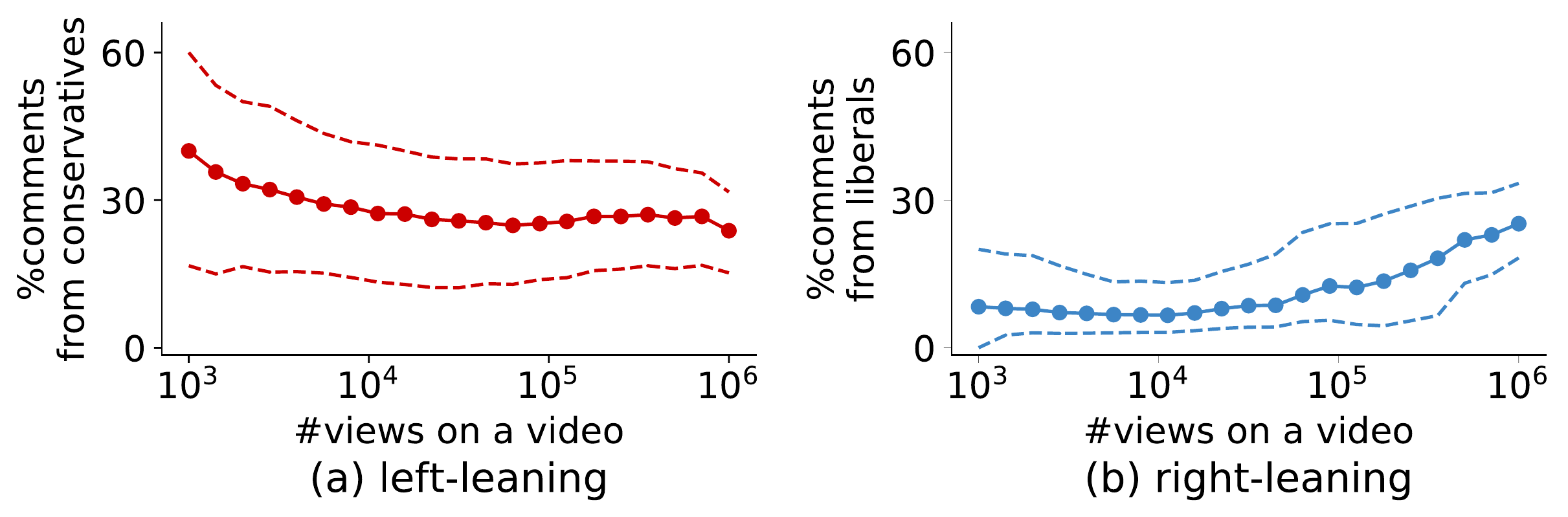}
	\caption{Analysis of videos: there were more cross-partisan comments on left-leaning videos than these on right-leaning videos.
	The x-axis shows the video view counts and it is divided into 21 equally wide bins in log scale.
	Each bin contains 553 to 7,527 videos.
	The lines indicate median and inter-quartile range.
	}
	\label{fig:measure-video}
\end{figure}

\subsection{How many cross-partisan comments do videos attract?}

We also quantified cross-partisan discussions on the video level.
For videos with at least 10 comments, we counted the number of comments posted by liberals and posted by conservatives.
\Cref{fig:measure-video} plots the fraction of cross-partisan comments on (a) left-leaning and (b) right-leaning videos.
We make two observations here:

First, higher fraction of cross-partisan comments occurred on left-leaning videos (median: 28\%, mean: 29.5\%) than on right-leaning videos (median: 8.6\%, mean: 13.4\%).
This result provides a new angle for explaining the creation of conservative echo chambers online~\cite{garrett2009echo,lima2018inside}.
For random viewers who watched right-leaning videos and browsed the discussions there, they would be exposed to very few comments from liberals.
On the other hand, users might experience relatively more balanced discussions on the left-leaning videos since about one in three comments there was made by conservatives.

Second, the correlations between video popularity and cross-partisan comments were opposite for left-leaning and right-leaning videos. When left-leaning videos attracted more views, the fraction of comments from conservatives became lower.
On the contrary, the fraction of comments from liberals was higher on right-leaning videos when the videos attracted more views.
This finding reveals potentially different strategies when politically polarized users carry out cross-partisan communication:
while conservatives occupy the discussion spaces in less popular left-leaning videos, liberals largely comment on high profile right-leaning videos.

\begin{figure}[!tbp]
    \centering
	\includegraphics[width=0.7\columnwidth]{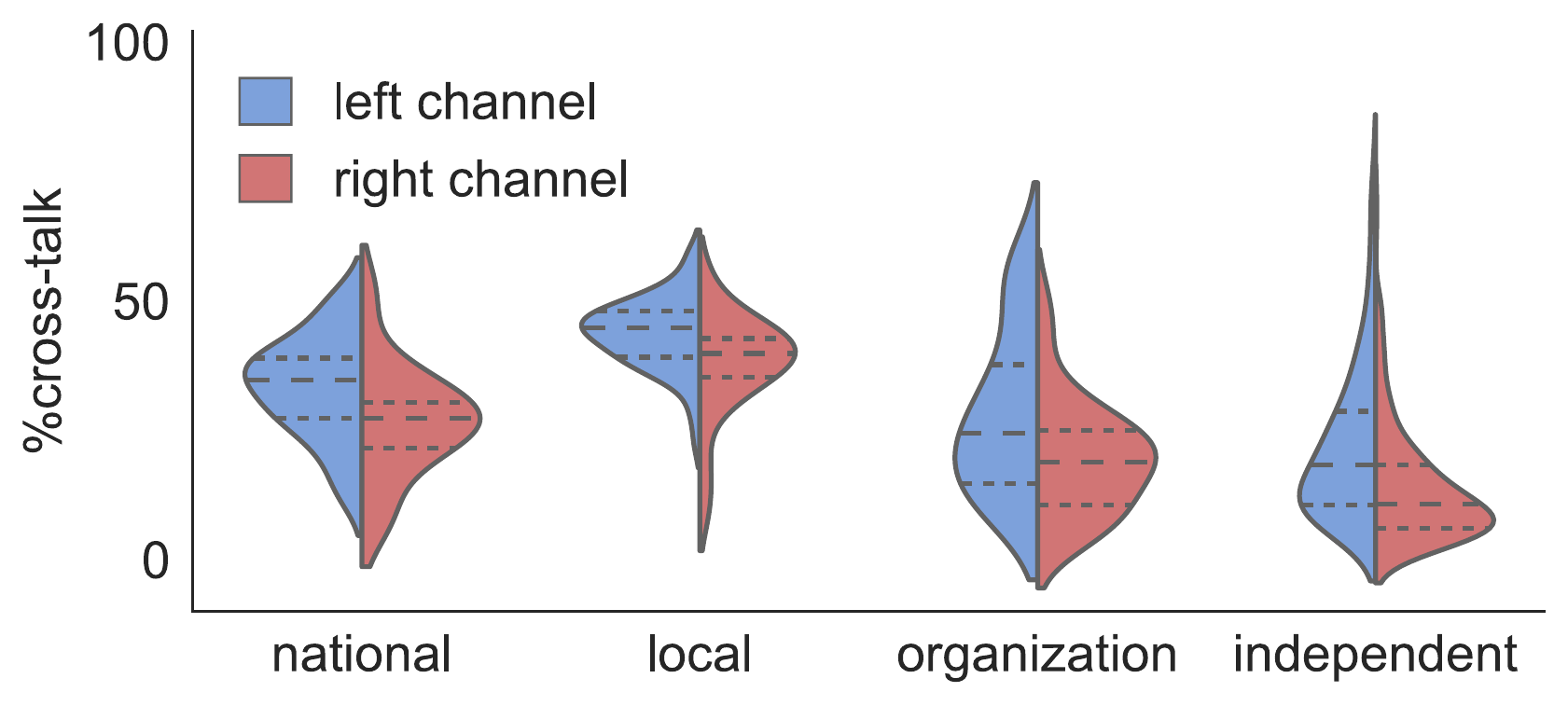}
	\caption{Analysis of media types. Local news channels attracted balanced audiences. By contrast, on right-leaning independent media, very few comments were posted by liberals. The outlines are kernel density estimates for the left-leaning and right-leaning channels.
	The center dashed line is the median, whereas the two outer lines denote the inter-quartile range.}
	\label{fig:measure-media}
\end{figure}

\subsection{Which media types attract more cross-partisan comments?}
Using our media coding scheme introduced in~\Cref{ssec:data-media}, we examined the extent of cross-talk in the four media types.
We excluded videos with less than 10 comments and then removed channels with less than five videos.
For each channel, we computed the mean fraction of cross-partisan comments over all of its videos, dubbed $\bar{\eta}(c)$.
The metric $\bar{\eta}(c)$ can be interpreted as the expected rate of cross-talk appearing on an average video of a given channel $c$.

\Cref{fig:measure-media} shows the distributions of $\bar{\eta}(c)$ in a violin plot, disaggregated by media types and political leanings.
Right-leaning channels received relatively fewer cross-partisan comments than the corresponding left-leaning channels across all four media types (statistically significant in one-sided Mann-Whitney U test at significance level of 0.05).
In particular, half of right-leaning independent media had fewer than 10.7\% comments from liberals, exposing their audience to a more homogeneous environment.
For example, ``Timcast'', who was the most commented right-leaning independent media in our dataset, had only 3.3 comments from liberals in every 100 comments.
This phenomenon stresses the potential harm for those who engage with ring-wing political commentary, because the discussions there often happen within the conservative echo chambers, which may in turn foster the circulation of rumors and misinformation~\cite{grinberg2019fake}.
On the other hand, the two prominent US news outlets -- CNN and Fox News -- were both crucial ground for cross-partisan discussions, having $\bar{\eta}(c)$ of 35.8\% and 31\%, respectively.

\section{Bias in YouTube's comment sorting}
\label{sec:system}

While biases in YouTube's video recommendation algorithms have been investigated~\cite{hussein2020measuring}, potential bias in its comment sorting algorithm is still unexplored. Presumably, YouTube ranks comments on the video pages based on recency, upvotes, and perhaps some reputation measure of the commenters, but not explicitly on whether the ideology of the commenter matches that of the channel. However, the suppression of cross-partisan communication may be a side effect of popularity bias~\cite{wu2019estimating}. For example, comments from the same ideological group may naturally attract more upvotes. 
Since YouTube has become an emerging platform for online political discussions to take place, it is important to understand the impacts of the comment ranking algorithm.

YouTube provides two options to sort the comments (see~\Cref{fig:data-webpage}a).
While the \texttt{Newest first} displays comments in a reverse-chronological order, \texttt{Top comments} is a black box algorithm that sorts and initially displays 20 root comments. The default view shows \texttt{Top comments}. 
In this section, we investigated the likelihood of position bias both (a) within the top 20 comments and (b) between the top 20 and the remaining comments.
We selected all videos with more than 20 root comments\footnote{For videos with no more than 20 root comments, the \texttt{Top comments} algorithm will show all their comments. Hence the display probability is 1 across all positions.}.
This yielded 64,876 left-leaning and 44,253 right-leaning videos.

\begin{figure}[tbp]
    \centering
	\includegraphics[width=0.99\columnwidth]{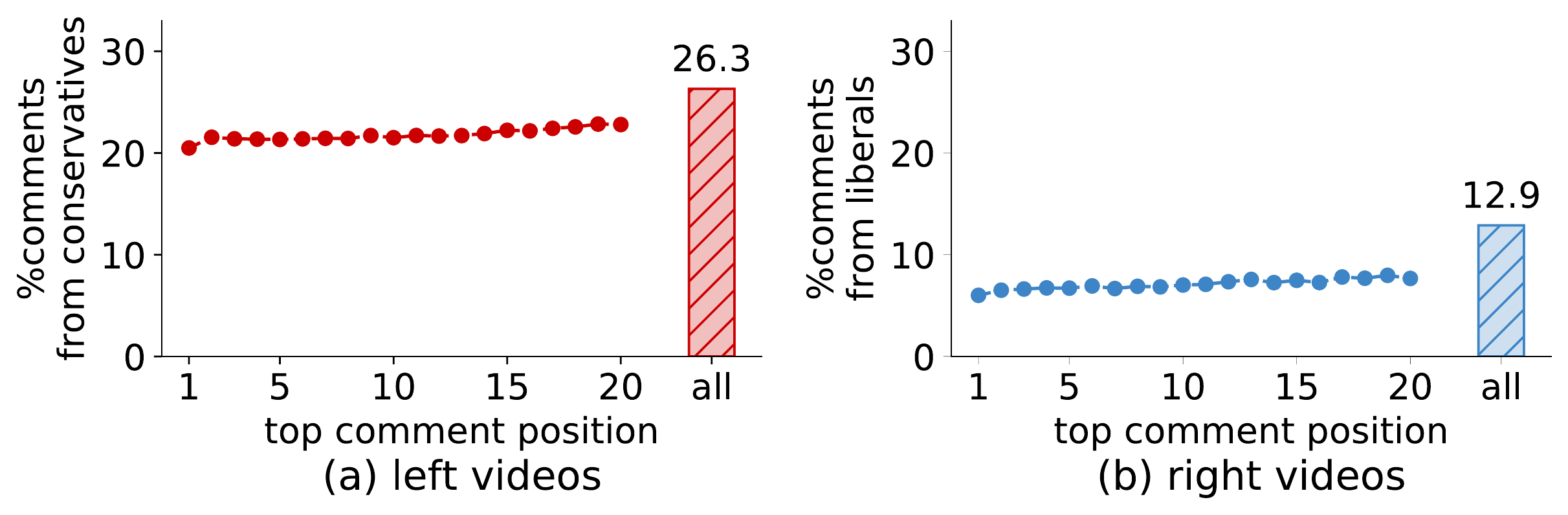}
	\caption{
	Root comments from conservatives on left-leaning videos and liberals on right-leaning videos were less likely to appear among the top 20 positions.
	The line charts show the fractions of cross-partisan comments at each position, while the bar charts show the overall fraction of cross-partisan comments.
	}
	\label{fig:measure-system-display}
\end{figure}

\Cref{fig:measure-system-display} shows the fraction of cross-partisan comments at each of the top 20 positions, as well as the overall prevalence over all sampled videos.
We observe a subtle trend over the top 20 successive positions.
More prominently, cross-partisan comments among the top 20 were less than the overall rate.
For example, for the 65K left-leaning videos, the fraction of comments from conservatives increased from 20.5\% at position 1 to 22.8\% at position 20.
However, overall 26.3\% of comments there were from conservatives. With more than 64K left videos and 44K right videos, the margins of error for all estimates were less than 0.41\%. Thus the comparisons between the top 20 positions and all positions were statistically significant.
This indicates that the sorting of \texttt{Top comments} creates a modest position bias against cross-partisan commentary, further diminishing the exposure of conservatives on left-leaning videos and liberals on right-leaning videos.

\section{Toxicity as a measure of quality}
\label{sec:quality}

While many political theorists have argued that exposure to diverse viewpoints has societal benefits, some researchers have also called attention to potential negative effects of cross-partisan communication.
\citet{bail2018exposure} showed that exposure to politically opposing views could increase polarization.
Even when the platforms can connect individuals with opposing views, the conversation may be more about shouting rather than meaningful discussions.
To this end, we complemented our prevalence analysis with a measure of the toxicity in the comments. Toxicity is a vague and subjective concept, operationalized as the fraction of human raters who would label something as ``a rude, disrespectful, or unreasonable comment that is likely to make you leave a discussion~\cite{wulczyn2017ex}''.

We obtained the toxicity scores for YouTube comments by querying the Perspective API~\cite{jigsaw2020}. 
The returned score ranges from 0 to 1, where scores closer to 1 indicate that a higher fraction of raters will label the comment as toxic.
We used 0.7 as a threshold as suggested in prior work~\cite{hua2020towards}.
For each cell of Table~\ref{table:toxic}\&\ref{table:toxic2}, we sampled 100K random comments that satisfy corresponding definitions, and then counted the fraction of comments deemed toxic.
Margins of error were less than 0.24\% with the sample size of 100K.
The data subsampling was to avoid making excessive API requests. 

Some YouTube comments are at the top level (i.e., root comments, \Cref{fig:data-webpage}b) and some are replies to other comments (\Cref{fig:data-webpage}c).
This brings a challenge of assigning comments to the correct targets.
Fortunately, YouTube employs the character ``@'' to specify the target (comment C in \Cref{fig:data-webpage}d is such case).
Hence we were able to curate two sets of comments: one contained root comments, the other contained replies to other users.

\begin{table}
    \centering
    \small
    \begin{tabular}{r|rr}
        \toprule
        \diagbox{from}{to} & left video & right video \\
        \midrule
        liberal & 15.73\% & \textbf{16.31\%} \\
        conservative & \textbf{15.77\%} & 14.44\%  \\
        \bottomrule
    \end{tabular}
    \caption{Percentage of root comments that are toxic. Bolded values are posts on opposite ideology videos. 
    }
    \label{table:toxic}
\end{table}

\begin{table}
    \centering
    \small
    \begin{tabular}{r|cc|cc}
        \toprule
        \multirow{2}{*}{\diagbox{from}{to}} & lib. & cons. & lib. & cons. \\
        & \multicolumn{2}{c}{on left video} & \multicolumn{2}{c}{on right video} \\
        \midrule
        liberal & 12.12\% & \textbf{18.24\%} & 13.42\% & \textbf{15.31\%} \\
        conservative & \textbf{15.24\%} & 11.11\% & \textbf{17.15\%} & 10.18\% \\
        \bottomrule
    \end{tabular}
    \caption{Percentage of replies that are toxic. Bolded values are replies between two users of opposite ideologies.
    }
    \label{table:toxic2}
\end{table}

\header{Results.}
\Cref{table:toxic} reports the frequency of toxic root comments.
We find that liberals and conservatives' root comments had about the same toxicity when posting on left-leaning videos.
However, conservatives posted fewer toxic root comments on right-leaning videos, and thus slightly fewer toxic root comments overall.

\Cref{table:toxic2} reports the frequency of toxic replies.
We find that replies to people of opposite ideology were much more frequently toxic, for both liberals and conservatives.
There also appears to be a ``defense of home territory'' phenomenon. Conservatives were significantly more toxic in their replies to liberals on right-leaning videos (17.15\%) than on left-leaning videos (15.24\%) and analogously for liberals responding to conservatives (18.24\%  on left-leaning vs. 15.31\% on right-leaning videos). Commenting on an opposing video generates more hostile responses than commenting on a same-ideology video. Interestingly, this holds true even for replies from people who share their ideology.
For example, liberals received more toxic replies from liberals on right-leaning videos (13.42\% toxic) than they did on left-leaning videos (12.12\% toxic).

\section{Privacy considerations}
All data that we gathered was publicly available on the YouTube website.
It did not require any limited access that would create an expectation of privacy such as the case of private Facebook groups. 
The analyses reported in this paper do not compromise any user identity.

For the public dataset that we are making available, however, there are additional privacy concerns due to new forms of search that become possible when public data is made available as a single collection.
In our released dataset of comments, we take efforts to make it difficult for someone who starts with a known comment authored by a particular YouTube user to be able to search for other comments written by the same user. Such a search is not currently possible with the YouTube website or API, and so enabling it would create a privacy reduction for YouTube users. To prevent this, we do not associate any user identifier, even a hashed one, with each comment text.

Without the ability to link comments by the authors, it will not be possible for other researchers to reproduce the training process for our HAN model that predicts a user's political leaning. The released dataset also does not associate predicted political leanings with YouTube user ids. The political leaning is a property predicted from the entire collection of comments by a user, and thus is not something that would be readily predicted from the user data that is easily available from YouTube. Furthermore, political leaning is considered sensitive personal information in some countries, including the European Union, which places restrictions on the collection of such information under the General Data Protection Regulation~\cite{gdpr2016regulation}. Unfortunately, this limits the ability of other researchers to produce new analyses of cross-partisan communication based on our released dataset.

Instead, we are releasing our trained HAN model. This will allow other researchers who independently collect a set of comments from a single user to predict the political leaning of that user.

\section{Discussion}
The results challenge, or at least complicate, several common narratives about the online political environment. The first is the echo chamber or filter bubble hypothesis, which posits that as people have more options of information sources and algorithm-assisted filtering of those sources, they will naturally end up with exposure only to ideologically-reinforcing information~\cite{carney2008secret}, as discussed in~\Cref{sec:related}.
We find that videos on left-leaning YouTube channels have a substantial number of comments posted by conservatives -- 26.2\%. While upvoting and the YouTube comment sorting algorithm somehow reduce their visibility, there are still more than 20\% top comments posted by conservatives. On right-leaning videos, there is somewhat less cross-talk, but even there on average nearly two of the top 20 comments are from liberals. Only on independent right-leaning channels such as ``Timcast'' do we see a comment stream that approaches an echo chamber with only conservatives posting. Liberals who are concerned about these channels serving as ideological echo chambers might do well to organize participation in them in an effort to diversify the ideologies that are represented.

Our results offer one direction for design exploration for platforms that want to reduce the extent to which their algorithms contribute to ideological echo chambers.
We have shown that user political leaning can be directly estimated by the textual features. Hence it is possible to directly incorporate political ideology into the comment sorting algorithm, not in order to filter out ideologically disagreeable content but to increase exposure to it. One simple approach would be to stratify sampling based on the fraction of cross-partisan comments, hence enforcing cross-talk to some extent. A more palatable approach might be to give a boost in the ranking only to those cross-partisan comments that receive a positive reaction in user upvoting.

The second narrative is that conservatives are less open to challenging political ideas, and more interested in staying in ideological echo chambers. On personality tests, conservatives score lower on openness to new experiences \cite{carney2008secret}. In a study of German speakers, openness to new experiences was modestly associated with consuming more news sources \cite{sindermann2020age}. In a study of Facebook users, those with higher openness tended to participate in a greater diversity of Facebook groups (i.e., groups with different central tendencies in the ideology of the participants) \cite{matz2021personal}.

To the contrary, we find that conservatives were far more likely to comment on left-leaning videos than liberals were to comment on right-leaning videos. More conservatives made at least one cross-partisan comment (82.3\% vs. 62.2\%) and the median fraction of cross-partisan comments among conservatives was 22.2\%, while the median for liberals was only 4.8\%. One possible interpretation is that some of the left-leaning channels that conservatives comment on are what are often considered as ``mainstream'' media, where people go to get news regardless of ideology. However, there is some evidence from other platforms that conservatives may, on average, seek out more counter-attitudinal information and interactions. On Facebook, \citet{bakshy2015exposure} found that conservatives clicked on more cross-partisan content than liberal did. On Twitter, \cite{grinberg2019fake} (see Figure 4) found that conservatives were more likely to retweet URLs from left-leaning (non-fake) news sources than liberals were to retweet URLs from right-leaning (non-fake) news sources. Also on Twitter, \cite{eady2019many} found that conservatives were more likely to follow media and political accounts classified as left-leaning than liberals were to follow right-leaning accounts.

An alternative narrative is that conservatives enjoy disrupting liberal conversations for the ``lulz'', that they initiate discord by eliciting strong emotional reactions~\cite{phillips2015we}. 
Given prior research on the general tone of cross-partisan communication~\cite{hua2020towards}, it is not surprising that we found cross-partisan comments were on average more toxic. 
We found, though, that conservatives were not significantly more toxic than liberals in their root comments on left-leaning channels, which casts some doubt on the idea that most of their interactions on left-leaning videos were trolling attempts. It is still possible that they were very strategic trolls, baiting liberals into toxic responses without themselves posting comments that people would judge as toxic. The conservative trolls story, however, is inconsistent with the finding that conservatives were less toxic in responses to liberals on left-leaning videos than they were on right-leaning videos. While there may be some conservative trolls who delight in disruption, there seems to be others who change their style to be more accommodating when they know that they are on the liberals' home territory. 

The reality is clearly more complicated than any of the existing simple narratives. Further theorizing is necessary to provide a coherent story of when and how liberals and conservatives tend to consume challenging information and engage in cross-partisan political interaction. To move in that direction, more interview and survey studies are needed to understand the motivations of why people participate in cross-partisan political online.

\section{Conclusion}
\label{sec:conclusion}
This paper presents the first large-scale measurement study of cross-partisan discussions between liberals and conservatives on YouTube.
We estimate the overall prevalence of cross-partisan discussions based on user political leanings predicted by a hierarchical attention model.
We find a large amount of cross-partisan commenting, but much more frequently by conservatives on left-leaning videos than by liberals on right-leaning videos.
YouTube’s comment sorting algorithm further diminishes the visibility of cross-partisan comments: they are somewhat less likely to appear among the top 20 positions. Even so, readers of comments on left-leaning videos are quite likely to encounter comments from conservatives and readers of comments on right-leaning videos from national and local media are likely to encounter some comments from liberals. Only on right-leaning independent media are the comments almost exclusively from conservatives. Lastly, we find that people tend to be slightly more toxic when they venture into channels with opposing ideologies, however they also receive much more toxic replies. The highest toxicity occurs when defending one's home territory -- liberals responding to conservatives on left-leaning videos and conservatives responding to liberals on right-leaning videos.

\section*{Acknowledgments}
This work is supported in part by the National Science Foundation under Grant No. IIS-1717688 and the AOARD project FA2386-20-1-4064.
We thank the Australia's National eResearch Collaboration Tools and Resources (Nectar) for providing computational resources.

\bibliography{youtube-crosstalk-ref}

\appendix

\section*{Appendix I}

\section{Percent of users who comment on both left-leaning and right-leaning channels}
\label{sec:si_cp_user}

\begin{figure}[h]
    \centering
	\includegraphics[width=0.99\columnwidth]{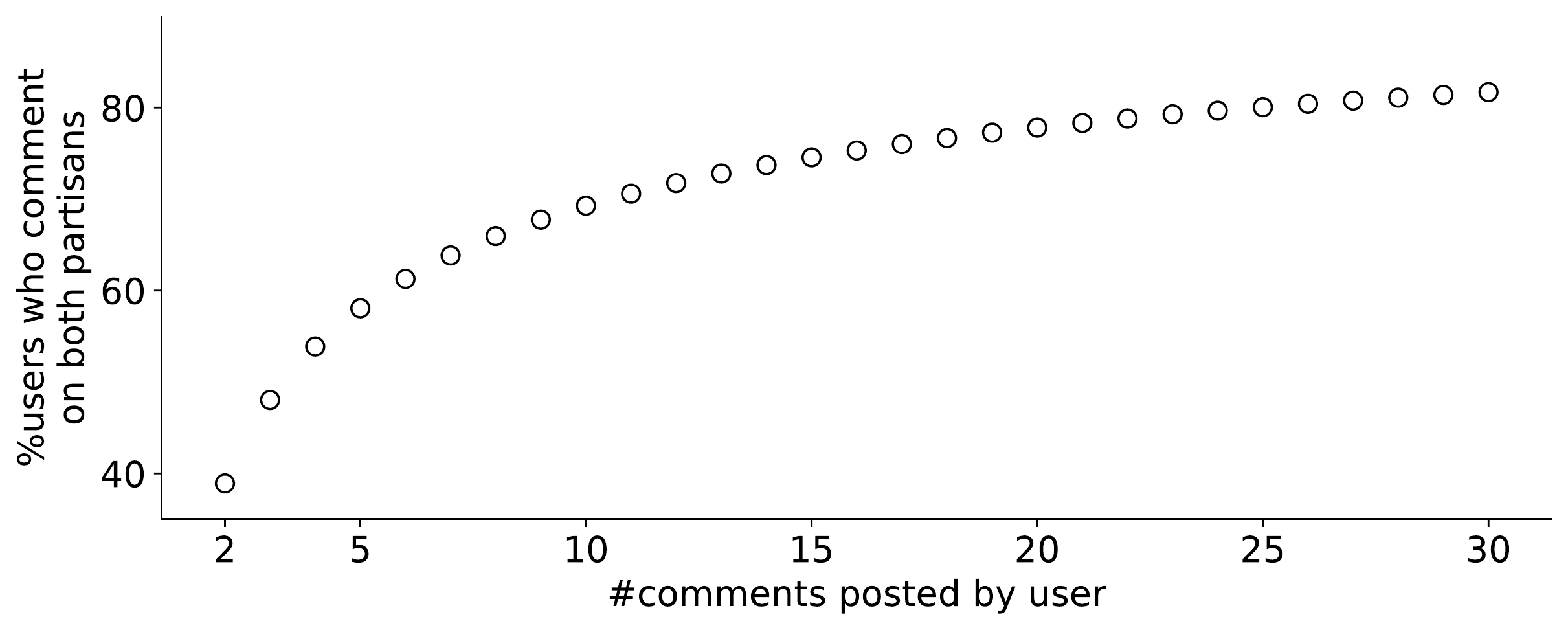}
	\caption{x-axis: minimal number of comments users posted; y-axis: portion of users who commented on both left-leaning and right-leaning channels.
	For users who posted at least 2 (or at least 25) comments, 39\% (or 80\%) of them commented on both partisan channels.}
\end{figure}

\end{document}